\documentclass[a4paper,twoside]{article}

\usepackage{epsfig}
\usepackage{subfigure}
\usepackage{calc}
\usepackage{amssymb}
\usepackage{amstext}
\usepackage{amsmath}
\usepackage{amsthm}
\usepackage{multicol}
\usepackage{pslatex}
\usepackage{apalike}
\usepackage{SCITEPRESS}
\usepackage[small]{caption}
\usepackage{listings}

\graphicspath{{./images/}}
\DeclareGraphicsExtensions{.pdf,.jpeg,.png,.jpg}

\begin{document}

\title{
Service Level Agreement Complexity  
\subtitle{Processing Concerns for Standalone and Aggregate SLAs}
}

\author{\authorname{Christopher C. Lamb\sup{1} and Gregory L. Heileman\sup{1}}
\affiliation{\sup{1}Department of Electrical and Computer Engineering, The University of New Mexico, Albuquerque, NM 87131}
\email{\{cclamb, heileman\}@ece.unm.edu}
}


\abstract{
\noindent In this paper, we examine the problem of a single provider offering multiple types of service level agreements, and the implications thereof.  In doing so, we propose a simple model for machine-readable service level agreements (SLAs) and outline specifically how these machine-readable SLAs can be constructed and injected into cloud infrastructures - important for next-generation cloud systems as well as customers.  We then computationally characterize the problem, establishing the importance of both verification and solution, showing that in the general case injecting policies into cloud infrastructure is NP-Complete, though the problem can be made more tractable by further constraining SLA representations and using approximation techniques.
}

\onecolumn \maketitle \normalsize \vfill

\section{Introduction}
The past few years have witnessed unprecedented expansion of commercial computing operations as the idea of cloud computing has become more mainstream and widely adopted by forward thinking technical organizational leadership.  This rate of adoption promises to increase in the near future as well.  With this expansion has come opportunity as well as risk, embodied by recent major service outages at leading cloud providers like Amazon.  These issues promise to become more difficult to control as managed infrastructure expands.  This expansion will simply not be possible without large amounts of automation in all aspects of cloud computing systems.

The current state of the art in cloud systems is poorly differentiated and not as customer-focused as it could be.  Current providers place the responsibility of monitoring performance and proving outages on the consumer \cite{ctrl:amazon-cloud-watch}.  Furthermore, providers as a whole usually provide one type of service level agreement (SLA) in a loosely-defined one-size-fits-all type of arrangement. Current SLAs are also very difficult to evaluate and manage \cite{Hilley-2009}.  This provides strong differentiating opportunities for smaller, second generation cloud system providers who have established the technology required to scalably manage multiple, competing SLAs on the same infrastructure in tandem with clear customer system visibility.

These second generation providers will rely on automated infrastructure management in order to scale.  One of the first steps toward automating these systems is automating SLA management and compliance.  Likewise, future cloud users are likely to build systems spanning multiple cloud providers, providing more difficult system management scenarios from their perspective as well \cite{ctrl:lamb-MCCCS}.

Herein, we elaborate the idea of applying usage management to a single system governed by multiple different types of SLAs.  We will define the problem, formally describe SLAs, analyze the implications of that formality with an eye toward efficient computability and the implications thereof.

In Section \ref{sec:cloud-models}, this paper begins by describing the different types of cloud computing models that generally exist today and how they manage services.  Immediately thereafter, we propose a possible future model in which users can have unique SLAs that more closely fit their needs rather than shoehorning their computing needs into a previously configured contract.  Then, in Section \ref{sec:SLA-defined}, we more formally define an SLA, and show how to convert one to an evaluatable expression.  In the following section we analyze the new SLA model and extract specific theoretical limits on computability and discuss implications thereof, showing solutions to SLAs in general to be NP-Complete.

\subsection{Previous Work}
As cloud computing is emerging as the future of utility systems hosting for consumer-facing applications.  In these kinds of systems, components, applications, and hardware are provided as utilities over the Internet with associated pricing schemes pegged by system demand.  Users accept specific Quality-of-Service (QoS) guidelines that providers use to provision and eventually allocate resources. These guidelines become the basis over which providers charge for services.

Over the past few years multiple service-based paradigms such as web services, cluster computing and grid computing have contributed to the development of what we now call cloud computing \cite{Bu:09}. Cloud computing distinctly differentiates itself from other service-based computing paradigms via a collective set of distinguishing characteristics:  market orientation, virtualization, dynamic provisioning of resources, and service composition via multiple service providers \cite{BuYeVeBrBr:09}. This implies that in cloud computing, a cloud-service consumer's data and applications reside inside that cloud provider's infrastructure for a finite amount of time.  Partitions of this data can in fact be handled by multiple cloud services, and these partitions may be stored, processed and routed through geographically distributed cloud infrastructures. These activities occur within a cloud, giving the cloud consumer an impression of a single virtual system.  These operational characteristics of cloud computing can raise concerns regarding the manner in which cloud consumer's data and applications are managed within a given cloud. Unlike other computing paradigms with a specific computing task focus, cloud systems enable cloud consumers to host entire applications on the cloud (i.e. software as a service) or to compose services from different providers to build a single system. As consumers aggressively start exploiting these advantages to transition IT services to external utility computing systems, the manner in which data and applications are handled within those systems by various cloud services will become a matter of serious concern \cite{Jamkhedkar:2010:IUM:1866870.1866885}.
\section{Cloud System Models}\label{sec:cloud-models}
\noindent Current cloud systems do not ignore SLA restrictions; rather, they are designed from the ground up to support a single type of SLA.  That SLA generally encompasses total system uptime and some kind of response time metric \cite{ctrl:amazon-sla,ctrl:rackspace-sla}.  If for some reason the cloud provider can no longer adhere to the terms outlined, some kind of compensation strategy generally applies to affected customers.  Future cloud providers can very well use the ability to support multiple SLAs as a way to differentiate available products from competitors.

\begin{figure}[!t]
\centering
\includegraphics[width=3in]{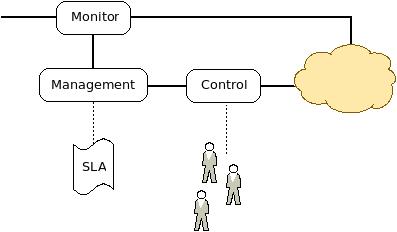}
\caption{Single SLA with Control Elements}
\label{fig:current-cloud-model}
\end{figure}

\subsection{Current Model}
Current systems like Amazon's EC2 or Rackspace products are designed around high availability, and this is reflected in the focus of their supplied SLAs.  This common design focus is also evident in the artifacts generated by other vendors \cite{ctrl:google-arch}.  Furthermore, Amazon offers clear guidance on how to develop systems that take advantage of their robust architecture as well as services that provide some measure of automatic scaling \cite{ctrl:amazon-best-practice,ctrl:amazon-fault-tolerant}.  This combination of market leading position and products and the extensive supplied guidance make Amazon a clear choice to examine when reflecting on the current state-of-the-art.

Amazon's Cloud Watch products used in tandem with Auto Scaling provide the ability to control the number of deployed instances in response to specific system loads, as shown in Figure \ref{fig:current-cloud-model} \cite{ctrl:amazon-cloud-watch,ctrl:amazon-auto-scale}.  Cloud Watch gives customers the ability to monitor various system performance metrics for their virtual machines, including but not limited to latency, processor use, and request counts.  Furthermore, users can set resource levels at which additional EC2 instances are created or destroyed.  This provides some level of personalized management and control over deployed systems within Amazon's cloud infrastructure.

\subsection{Future Reference Model}
\begin{figure}[!t]
\centering
\includegraphics[width=3in]{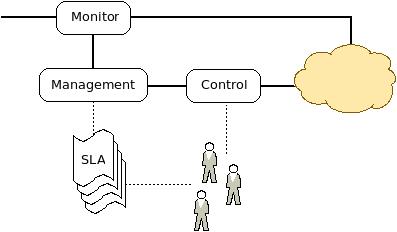}
\caption{Multiple SLA Architectural Integration}
\label{fig:future-cloud-model}
\end{figure}

While current cloud service providers focus on a single QoS metric, future providers may very well begin to provide multiple metrics over which they will define service levels, as shown in Figure \ref{fig:future-cloud-model}.  This is not without precedent --- just as airlines provide the same essential product at different service points, cloud providers could supply system hosting via disparate service levels, including divergent service metric definitions.  For example, current architectures support uptime and availability as the primary managed metric from an SLA perspective.  Future architectures could support uptime and availability, as well as specific latency, bandwidth, and geo-location sensitive hosting parameters.  These kinds of SLAs would also continue to outline penalties when any of the conditions of that SLA were violated.  Unlike current SLAs however, these could also differentiate based on the magnitude of the imposed penalty, with different classifications of service mapping to increasingly large penalties on service failure.

While industry does seem to certainly be trending in this direction, as indicated by the development of tools supporting user-centric infrastructure monitoring and management, this kind of control is not yet embedded into contracts of any kind, much less agreements that are machine-readable. Furthermore, this kind of management is still manual and cannot scale to the levels needed to manage Internet-scale systems.

\section{Service Level Agreements Defined}\label{sec:SLA-defined}
As we have seen, SLAs generally consist of a set of conditions of use under which the SLA is binding, a set of obligations that the provider will adhere to if the customer adheres to the set conditions, and two sets of penalties, one penalizing the provider when breaching obligations, and another penalizing the customer when breaching conditions of use.  Conditions are generally loosely defined, while provider obligations are much more rigorously constructed.  Generally however, conditions and obligations in this context can be viewed as defined by {\it objectives} which are measured by {\it indicators}.  In the case of provider-centric obligations, these are commonly defined as service level indicators (SLIs) and objectives (SLOs).

With this general understanding of SLAs and related SLIs and SLOs, we can create a non-specific definition of an SLA as a list of tuples related via boolean operators including $($, $)$, $\lor$, $\land$, and $\lnot$.  Each tuple is a clause in the SLA:

\begin{eqnarray}
CLAUSE = \lbrace (I,O,E,P) \rbrace, \\
OPERATOR = \lbrace ( , ) , \lor, \land, \lnot \rbrace \\
SLA = \lbrace ( o_{0}, c_{0}, o_{1}, c_{1}, ..., c_{n}, o_{n + 1} ) \rbrace , \\ c \in CLAUSE, o \in OPERATOR \notag
\end{eqnarray}

Where $ I $ is an indicator function, $ \forall i \in I, i : () \rightarrow \tau $, which retrieve indicator values, generally related to some kind of SLI or customer condition indicator.  Parameter $ O $ is a set of values derived from SLOs or customer condition objectives such that $ \forall o \in O, o : P(\tau) $.  Parameter $ E $ is a set of predicates that indicate whether a specific indicator complies with its objectives, where $ \forall e \in E, e : ( () \rightarrow \tau ) \times P(\tau) \rightarrow bool $.  Parameter $ P $ is a set of penalty functions, $ \forall p \in P, p : Z \rightarrow Z $, where the first argument is generally an elapsed time value.

For example, say we are a customer of Nimbus Cloud Corporation, and we have an SLA in which Nimbus provides guaranteed 100\% uptime and packet latency between 300 and 650 milliseconds.  Nimbus does not have any customer conditions specified within its SLAs.  This then gives us a machine evaluateable SLA:
\begin{align}
SLA_{nimbus} = ( ( uptime\_monitor() : bool, \notag\\
\lbrace true \rbrace, \notag \\
uptime\_evaluator( monitor : () \rightarrow bool, \lbrace true \rbrace) : bool, \notag \\
uptime\_penalty\_evaluator( T : Z ) : Z ), \notag \\
\land, \notag \\
( latency\_monitor() : Z, \notag \\
\lbrace 300, 650 \rbrace, \notag \\
latency\_evaluator( monitor : () \rightarrow Z, \lbrace 300, 650 \rbrace) : bool, \notag \\
latency\_penalty\_evaluator( T : Z ) : Z) ) \notag
\end{align}
This more rigorous SLA allows users to monitor obligations and determine penalties when triggered.  Furthermore, when structuring SLAs in this way, we are able to combine any single SLA with other SLAs but extracting the clauses and creating a new aggregate SLA, which is just the conjunction of the combined SLAs, such that for $ A, B, C \in SLA $, $ ( A \land B \land C ) \in SLA $.  We will further define SLA-SAT as the problem of establishing whether a given SLA or group of SLAs have a solution.

Also note that evaluators, as they compare the output of a monitor to an input value, can run in constant time.  Monitors, as they retrieve specific contextual values from a running system, execute over some assumed constant time not influenced by the SLA.

\section{Evaluating and Verifying Service Level Agreements}\label{sec:SLA-analysis}
Now that we have rigorously defined our SLAs, notice that the SLA evaluation functions are predicates, and can be curried for later execution if needed.  This allows us to begin a more fundamental analysis of SLAs and their capabilities.

\subsection{Computational and Space Complexity}
In Section \ref{sec:SLA-defined}, we defined an SLA to essentially be a sequence of evaluatable predicates.  These evaluatable predicates are related in some way; currently, an SLA is the conjunction of these predicates.  As these predicates can be created prior to evaluation, and at evaluation time require no specific arguments once appropriately curried, we can define these predicates as boolean {\it terms}.  Ergo, once we have created a group of predicates and transformed them into terms, we are evaluating an arbitrary boolean equation - in other words, we are verifying an instance of the Boolean Satisfiability Problem, or SAT.\\

\noindent {\bf Claim 1:} SLA-SAT is in {\bf NP}.\\

\noindent {\bf Proof:} The following verifier runs in polynomial time in the length of an SLA-SAT.\\

\noindent {\sl Input}: $ \sigma \in $ SLA-SAT, a verifier $v_{SAT}$ for SAT. \\
\noindent {\sl Output}: TRUE if the SLA is satisfied FALSE otherwise.
\begin{tabbing}
~~~~\=~~~~\=~~~~\=~~~~\= \\
\noindent Verifier $V_{SLA-SAT} \langle \sigma, v_{SAT}  \rangle$ : \\
\> $ l \leftarrow $ list \\
\> \underline{for} each element $ e \in \sigma $ :\\
\>\> \underline{if} e is a clause : \\
\>\>\> extract monitor $ m \in M $ from $ c $\\
\>\>\> extract objective $ o \in O $ from $ c $\\
\>\>\> extract evaluator $ v \in E $ from $ c $\\
\>\>\> $ r \leftarrow v ( m, o ) $\\
\>\>\> add $ r $ to $ l $ \\
\>\> \underline{else} : \\
\>\>\> add $ e $ to $ l $ \\
\>\> \underline{endif} \\
\> \underline{endfor} \\
\> \underline{return} $ v_{SAT}(l) $ \\
\end{tabbing}

\noindent We can furthermore establish that SLA-SAT is NP-Hard, establishing NP-Completeness.\\

\noindent {\bf Claim 2:} SAT $ \leq_{p} $ SLA-SAT.\\

\noindent {\bf Proof:} Define a function $ f : ( \beta ) \rightarrow \sigma $, where $ \beta $ is a generalized boolean formula consisting of boolean clauses related by the logical operators $($, $)$, $\land$, $\lor$, and $\lnot$ and $ \sigma \in $ SLA-SAT. $ f $ will parse through $ \beta $, skipping operators, and transforming boolean elements $ \zeta $ into tuples $ t \in CLAUSE $ such that $ t = ( null, null, f : ( () \rightarrow \tau \times P(\tau) \rightarrow bool ) \rightarrow \zeta, null ) $.  $ f $ then adds the operator or transfurmed boolean element to a list.  When $ f $ reaches the end of the boolean formula, it returns the list. \\

\noindent We can also show the inverse.\\

\noindent {\bf Claim 3:} SLA-SAT $ \leq_{p} $ SAT.\\

\noindent {\bf Proof:} Define a function $ g : ( \sigma ) \rightarrow \beta $, where $ \beta $ is a generalized boolean formula consisting of boolean clauses related by the logical operators $($, $)$, $\land$, $\lor$, and $\lnot$ and $ \sigma \in $ SLA-SAT.  Evaluate each element within $\sigma$, copying each operator into a boolean expression $\beta$, and transforming each clause by evaluating $e \in E$ with contained indicators and objectives $i \in I$ and $o \in O$.  Copy the resulting boolean value into the expression $\beta$.  When finished traversing $\sigma$, return $\beta$. \\

Therefore, as SAT is NP-Complete, and provably difficult to solve, SLA-SAT is NP-Complete as well \cite{comptheory:sipser:intro-comp-theory}.  However, 3SAT, a subset of SAT, is equally difficult, while 2SAT is not.  2SAT is firmly in the computational class P; in fact, 2SAT is NL-Complete as well, so we know it is solvable in an amount of space logarithmic in the number of boolean terms \cite{comptheory:papadimitriou:computational-complexity}; it is widely believed that both SAT and 3SAT cannot be solved in logarithmic or less space.  Finally, as 2SAT is NL-Complete, we also know it is contained within NC$^{2}$, and as such is highly parallelizeable \cite{comptheory:papadimitriou:computational-complexity}.  This implies that SLAs should be implemented or aggregated in a form no more complex than 2SAT to facilitate efficient processing.

\subsection{Verification v. Solution}
General SLA use focuses on verification rather than solution.  That is to say, with respect to a given SLA, both the user and provider is more concerned with whether the system is currently compliant with all SLA terms.  In future use, this may very well no longer be the case.  For example, imagine a cloud system from the user's perspective that spans multiple cloud providers; a single general purpose provider for general computing, data storage, and queuing, a specific-use provider for an unique set of algorithms of some kind (say, market modeling algorithms), and finally a Content Delivery Network (CDN) provider.  Each of these providers have a unique SLA with multiple conditions.  In this particular case, the user may need to know if the given system can work together at all under the terms of the SLAs, and if so, under what conditions.  As the user has combined all the SLAs composing the system and is attempting to find some combination of terms that satisfies the resulting boolean formula, the user is in essence attempting to find a solution for this instance of $ SAT $, a known NP-Complete problem.  Likewise, a cloud provider may need to solve similar problems where the SLAs at issue are the whole of SLAs issued to the entire provider's customer base.

Nontrivial generalized SLAs may be too difficult to solve effectively without using some kind of approximate heuristic or $ SAT $ Solver \cite{Hochbaum:1996:AAN:241938,ctrl:satcompetition}.  If these SLAs are formulated in at most a $ 2SAT $ style problem however, they are suddenly much more tractable, easier to work with, and amenable to efficient solution.  Keep in mind, even generalized SLAs can be efficiently verified.

\section{Conclusions and Future Works}
\noindent Herein, we started by going over the current generalized state of most cloud systems from an SLA perspective, differentiating between current architectures that incorporate SLA ideas into the design itself with possible futures architectures that incorporate pluggable SLAs with varying indicators and objectives.  We then generalized SLAs into sets of quadruples containing a monitoring function, a set of values defining acceptable ranges returned from the monitoring functions, an evaluation function, and a penalty evaluating function, and demonstrated how this formulation could be used with a specific example.  With this in place, we then demonstrated that the generalized SLA problem is equivalent to SAT, and therefore is NP-Complete.  We finally covered the implications and theoretical limits implied by this NP-Completeness, validating the applicability of this work by designing a realistic control model using these ideas.

This is preliminary work into establishing the theoretical bounds surrounding effective automated control of cloud systems within Internet-scale systems.  Furthermore, the SLA modeled was fairly simple, and only took into account externally-evaluatable metrics in a black-box arrangement.  SLAs can very well outline other operational parameters, like specific data routing, fine-grained usage management, or encryption requirements.  Likewise, experimental evidence supporting these control ideas will be vital to promoting acceptance and action around these concepts within both cloud service provider systems and user configured applications.

\bibliographystyle{apalike}
{\small
\bibliography{bib/emr,bib/drm,bib/ctrl,bib/readings}}

\begin{thebibliography}{}

\bibitem[ctr, 2011a]{ctrl:amazon-cloud-watch}
 (2011a).
\newblock {A}mazon {C}loudwatch.
\newblock {http://aws.amazon.com/cloudwatch/}.

\bibitem[ctr, 2011b]{ctrl:amazon-sla}
 (2011b).
\newblock {A}mazon {EC2} {SLA}.
\newblock {http://aws.amazon.com/ec2-sla/}.

\bibitem[ctr, 2011c]{ctrl:amazon-auto-scale}
 (2011c).
\newblock {{A}uto {S}caling}.
\newblock {http://aws.amazon.com/autoscaling/}.

\bibitem[ctr, 2011d]{ctrl:rackspace-sla}
 (2011d).
\newblock {R}ackspace {C}loud - {SLA}.
\newblock {http://www.rackspace.com/cloud/legal/sla/}.

\bibitem[ctr, 2011e]{ctrl:satcompetition}
 (2011e).
\newblock Sat competitions.
\newblock {http://www.satcompetition.org/}.

\bibitem[Barr et~al., 2010]{ctrl:amazon-fault-tolerant}
Barr, J., Narin, A., and Varia, J. (2010).
\newblock {B}uilding {F}ault-{T}olerant {A}pplications on {AWS}.

\bibitem[Buyya, 2009]{Bu:09}
Buyya, R. (2009).
\newblock Market-oriented cloud computing: Vision, hype, and reality of
  delivering computing as the 5th utility.
\newblock In {\em Proceedings of the 2009 9th IEEE/ACM International Symposium
  on Cluster Computing and the Grid}, CCGRID '09, pages 1--, Washington, DC,
  USA. IEEE Computer Society.

\bibitem[Buyya et~al., 2009]{BuYeVeBrBr:09}
Buyya, R., Yeo, C.~S., Venugopal, S., Broberg, J., and Brandic, I. (2009).
\newblock Cloud computing and emerging {IT} platforms: {V}ision, hype, and
  reality for delivering computing as the 5th utility.
\newblock {\em Future Generation Computer Systems}, 25(6):599--616.

\bibitem[Dean, 2009]{ctrl:google-arch}
Dean, J. (2009).
\newblock {D}esigns, {L}essons and {A}dvice from {B}uilding {L}arge
  {D}istributed {S}ystems.
\newblock Presented at the Large-Scale Distributed Systems and Middleware
  (LADIS) Conference.

\bibitem[Hilley, 2009]{Hilley-2009}
Hilley, D. (2009).
\newblock Cloud computing: A taxonomy of platform and infrastructure-level
  offerings.
\newblock Technical report, Georgia Institute of Technology.

\bibitem[Hochbaum, 1997]{Hochbaum:1996:AAN:241938}
Hochbaum, D.~S., editor (1997).
\newblock {\em Approximation algorithms for NP-hard problems}.
\newblock PWS Publishing Co., Boston, MA, USA.

\bibitem[Jamkhedkar et~al., 2010]{Jamkhedkar:2010:IUM:1866870.1866885}
Jamkhedkar, P.~A., Heileman, G.~L., and Lamb, C.~C. (2010).
\newblock An interoperable usage management framework.
\newblock In {\em Proceedings of the tenth annual ACM workshop on Digital
  rights management}, DRM '10, pages 73--88, New York, NY, USA. ACM.

\bibitem[Lamb et~al., 2011]{ctrl:lamb-MCCCS}
Lamb, C.~C., Jamkhedkar, P.~A., Heileman, G.~L., and Abdallah, C.~T. (2011).
\newblock Managed control of composite cloud systems.
\newblock In {\em 6th IEEE International Conference on System of Systems
  Engineering (SOSE)}. IEEE.

\bibitem[Papadimitriou,
  1994]{comptheory:papadimitriou:computational-complexity}
Papadimitriou, C.~M. (1994).
\newblock {\em {Computational complexity}}.
\newblock Addison-Wesley, Reading, Massachusetts.

\bibitem[Sipser, 1997]{comptheory:sipser:intro-comp-theory}
Sipser, M. (1997).
\newblock {\em Introduction to the Theory of Computation}.
\newblock International Thomson Publishing, 1st edition.

\bibitem[Varia, 2010]{ctrl:amazon-best-practice}
Varia, J. (2010).
\newblock {A}rchitecting for the {C}loud: {B}est {P}ractices.

\end{thebibliography}

\end{document}